\documentclass{emulateapj}
\usepackage{apjfonts}
\usepackage{graphicx}
\usepackage{xspace}
\usepackage{natbib}

\newcommand{\Msun}      {\mbox{$\rm\,M_{\mathord\odot}$}}

\begin{document}

\submitted{Accepted by ApJ June 2016}

\def\lsim{\mathrel{\lower .85ex\hbox{\rlap{$\sim$}\raise
.95ex\hbox{$<$} }}}
\def\gsim{\mathrel{\lower .80ex\hbox{\rlap{$\sim$}\raise
.90ex\hbox{$>$} }}}

\title{A \emph{NuSTAR} observation of the reflection spectrum of the low mass X-ray binary 4U 1728-34}

\author{Clio C. Sleator\altaffilmark{1},
John A. Tomsick\altaffilmark{1},
Ashley L. King\altaffilmark{2},
Jon M. Miller\altaffilmark{3},
Steven E. Boggs\altaffilmark{1},
Matteo Bachetti\altaffilmark{4,5},
Didier Barret\altaffilmark{4,5},
J{\'e}r{\^o}me Chenevez\altaffilmark{6},
Finn E. Christensen \altaffilmark{6},
William W. Craig\altaffilmark{7,8},
Charles J. Hailey\altaffilmark{8},
Fiona A. Harrison\altaffilmark{9},
Farid Rahoui\altaffilmark{10,11},
Daniel K. Stern\altaffilmark{12},
Dominic J. Walton\altaffilmark{9},
William W. Zhang\altaffilmark{13}}

\altaffiltext{1}{Space Sciences Laboratory, 7 Gauss Way, University of California, 
Berkeley, CA 94720-7450, USA}
\altaffiltext{2}{KIPAC, Stanford University, 452 Lomita Mall, Stanford, CA 94305, USA}
\altaffiltext{3}{Department of Astronomy, The University of Michigan, 500 Church Street, Ann Arbor, MI 48109-1046, USA}
\altaffiltext{4}{Universite de Toulouse, UPS-OMP, Toulouse, France}
\altaffiltext{5}{CNRS, Institut de Recherche en Astrophysique et Planetologie, 9 Av. colonel Roche, BP 44346, F-31028, Toulouse cedex 4, France}
\altaffiltext{6}{DTU Space, Technical University of Denmark, Elektrovej 327-328, Lyngby, DK}
\altaffiltext{7}{Lawrence Livermore National Laboratory, Livermore, CA}
\altaffiltext{8}{Columbia Astrophysics Laboratory and Department of Astronomy, Columbia University, 550 West 120th Street, New York, NY 10027, USA}
\altaffiltext{9}{Cahill Center for Astronomy and Astrophysics, California Institute of Technology, Pasadena, CA 91125, USA}
\altaffiltext{10}{European Southern Observatory, Karl Schwarzschild-Strasse 2, D-85748 Garching bei MŸnchen, Germany}
\altaffiltext{11}{Department of Astronomy, Harvard University, 60 Garden Street, Cambridge, MA 02138, USA}
\altaffiltext{12}{Jet Propulsion Laboratory, California Institute of Technology, 4800 Oak Grove Drive, Pasadena, CA 91109, USA}
\altaffiltext{13}{NASA Goddard Space Flight Center, Greenbelt, MD 20771, USA}

\begin{abstract}
We report on a simultaneous \emph{NuSTAR} and \emph{Swift} observation of the neutron star low-mass X-ray binary 4U 1728-34. We identified and removed four Type I X-ray bursts during the observation in order to study the persistent emission. The continuum spectrum is hard and well described by a black body with $kT=$ 1.5 keV and a cutoff power law with $\Gamma=$ 1.5 and a cutoff temperature of 25 keV. Residuals between 6 and 8 keV provide strong evidence of a broad Fe K$\alpha$ line. By modeling the spectrum with a relativistically blurred reflection model, we find an upper limit for the inner disk radius of $R_{\rm in}\leq2 R_{\rm ISCO}$. Consequently we find that $R_{\rm NS}\leq23$ km, assuming $M=1.4\Msun$ and $a=0.15$. We also find an upper limit on the magnetic field of $B\leq2\times10^8$ G.
\end{abstract}


\section{Introduction}

Iron emission lines with energies from 6.4 to 7.1 keV have been detected in some neutron star X-ray systems (e.g. \citealt{gottwald95,miller13,degenaar15,king15}). These lines are most likely due to the fluorescent K$\alpha$ transition of iron, caused by the reflection of hard X-ray emission onto an area of the accretion disk close to the compact object \citep{fabian89}. Relativistic and Doppler effects distort the profile of the line, broadening it significantly and skewing it to low energies \citep{miller07,reynolds03}. From this unique shape, one can measure interesting properties of the accretion disk, including its inner radius. An upper limit for the neutron star radius can thus be inferred from the reflection spectrum, which is critical to understanding the neutron star equation of state \citep{lattimer07,cackett10}.

4U 1728-34 is a neutron star low-mass X-ray binary (LMXB) of the atoll class \citep{lewin76, hasinger89}, with an estimated distance of 4.1-5.1 kpc \citep{disalvo00,galloway03}. It exhibits Type-1 X-ray bursts caused by thermonuclear burning on the stellar surface \citep{galloway10}. From burst oscillations, \cite{strohmayer96} measured a spin frequency of $363\pm6$ Hz.

A broad emission line at 6.7 keV has been detected in several spectral analyses of 4U 1728-34 performed with satellites such as \emph{XMM-Newton} \citep{ng10,egron11}, \emph{INTEGRAL} \citep{tarana11}, \emph{RXTE} \citep{piraino00,seifina11} and \emph{BeppoSAX} \citep{disalvo00,piraino00,seifina11}. The continuum spectrum is generally composed of a soft component described as thermal emission from the stellar surface or accretion disk, and a hard component described as Comptonization.

The \emph{Nuclear Spectroscopic Telescope Array} (\emph{NuSTAR}; \citealt{harrison13}) has been successful in revealing iron K$\alpha$ lines and reflection spectra in neutron stars (e.g. \citealt{miller13}, \citealt{degenaar15}, \citealt{king15}). In this work we analyze a coordinated \emph{NuSTAR} and \emph{Swift} observation of 4U 1728-34 in an effort to further constrain its reflection spectrum and thus the neutron star radius.

\section{Observations and Data Reduction}

\begin{figure}
\includegraphics[width=0.5\textwidth]{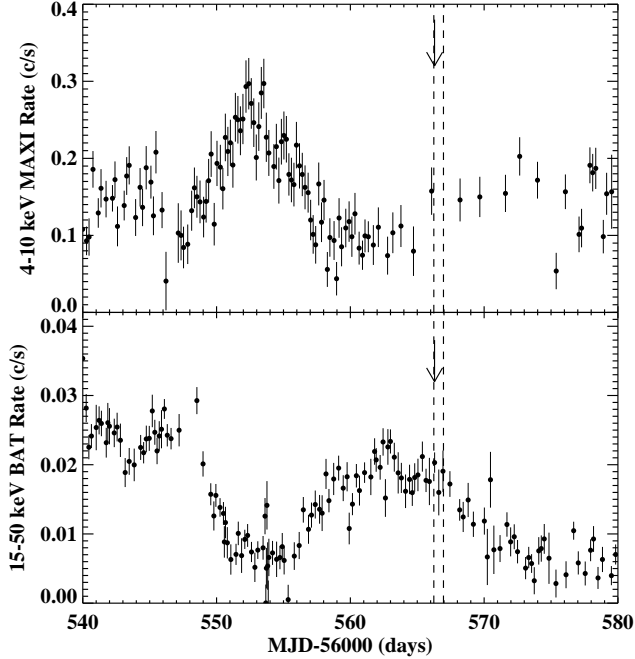}
\caption{\emph{MAXI} and \emph{BAT} light curves with the time of the \emph{NuSTAR} observations marked by the dashed lines. The arrow marks the time of the \emph{Swift} observation.\label{fig:maxibat}}
\end{figure}

\emph{NuSTAR} observed 4U 1728-34 on 2013 October 1 for 33.5 ks (Obs ID 80001012002; Figure \ref{fig:maxibat}). The data were processed with the \emph{NuSTAR} Data Analysis Software ({\ttfamily NuSTARDAS}) version 1.4.1 and {\ttfamily CALDB} 20150702, resulting in 27 ks of net exposure time. We extracted spectra from the FPMA and FPMB focal planes using a circular region with a $180\arcsec$ radius centered at the source position. We used a $671\arcsec\times114\arcsec$ rectangular background region centered $388\arcsec$ away from the source position. At 5 keV, the ratio of source count rate to background count rate is 200, and, at 50 keV, this ratio is 3, indicating that the spectrum is not very sensitive to the background estimate. Due to a known shift in gain offset that has not been properly accounted for in the current response files used for the reduction in this paper, we restrict our analysis to the 4.5-78.4 keV band\footnote{Confirmed by \emph{NuSTAR} instrument team (Kristin Madsen, private communication)}.

An observation (ObsID 00080185001) of 4U~1728--34 was made with the {\em Swift} \citep{gehrels04} X-ray telescope (XRT; \citealt{burrows05}) near the beginning of the {\em NuSTAR} observation (see Figures \ref{fig:maxibat} and \ref{fig:hardness}). The XRT was operated in Windowed Timing mode to avoid photon pile-up. Although the XRT observation lasted for two {\em Swift} orbits, the source was near the edge of the active area of the detector for the first orbit, and we produced an energy spectrum using only the second orbit, giving an exposure time of 1068\,s.  As this source has a relatively high column density, we used only grade 0 photons as this is the recommended procedure for high column density sources\footnote{see http://www.swift.ac.uk/analysis/xrt/digest\_cal.php\#abs}. The 0.7--10\,keV spectrum was extracted from a circular region with a 20 pixel ($47^{\prime\prime}$) radius, and background was taken from a region far from the source.  We obtained a source count rate in the 0.7--10\,keV band of 17.7\,c/s.  For spectral fitting, we calculated a response matrix appropriate for grade 0 photons (file swxwt0s6\_20130101v015.rmf) and used {\ttfamily xrtmkarf} with an exposure map to produce the ancillary response file.

All spectra were analyzed using XSPEC version 12.8 \citep{xspec}. All fits were made assuming \cite{vern} cross sections and \cite{wilm} abundances. The spectra were binned such that the signal-to-noise ratio in each bin is $15\sigma$. To better constrain the low energy spectrum, particularly the column density, we fit the \emph{Swift} spectrum together with the \emph{NuSTAR} spectra. Due to flux variations between the instruments, we added a multiplicative constant in each fit. We fixed the constant for the \emph{NuSTAR} FPMA spectrum to unity and allowed the constants for the \emph{NuSTAR} FPMB and \emph{Swift} spectra to vary.

\section{Analysis and Results}

\begin{figure}
\includegraphics*[width=0.5\textwidth]{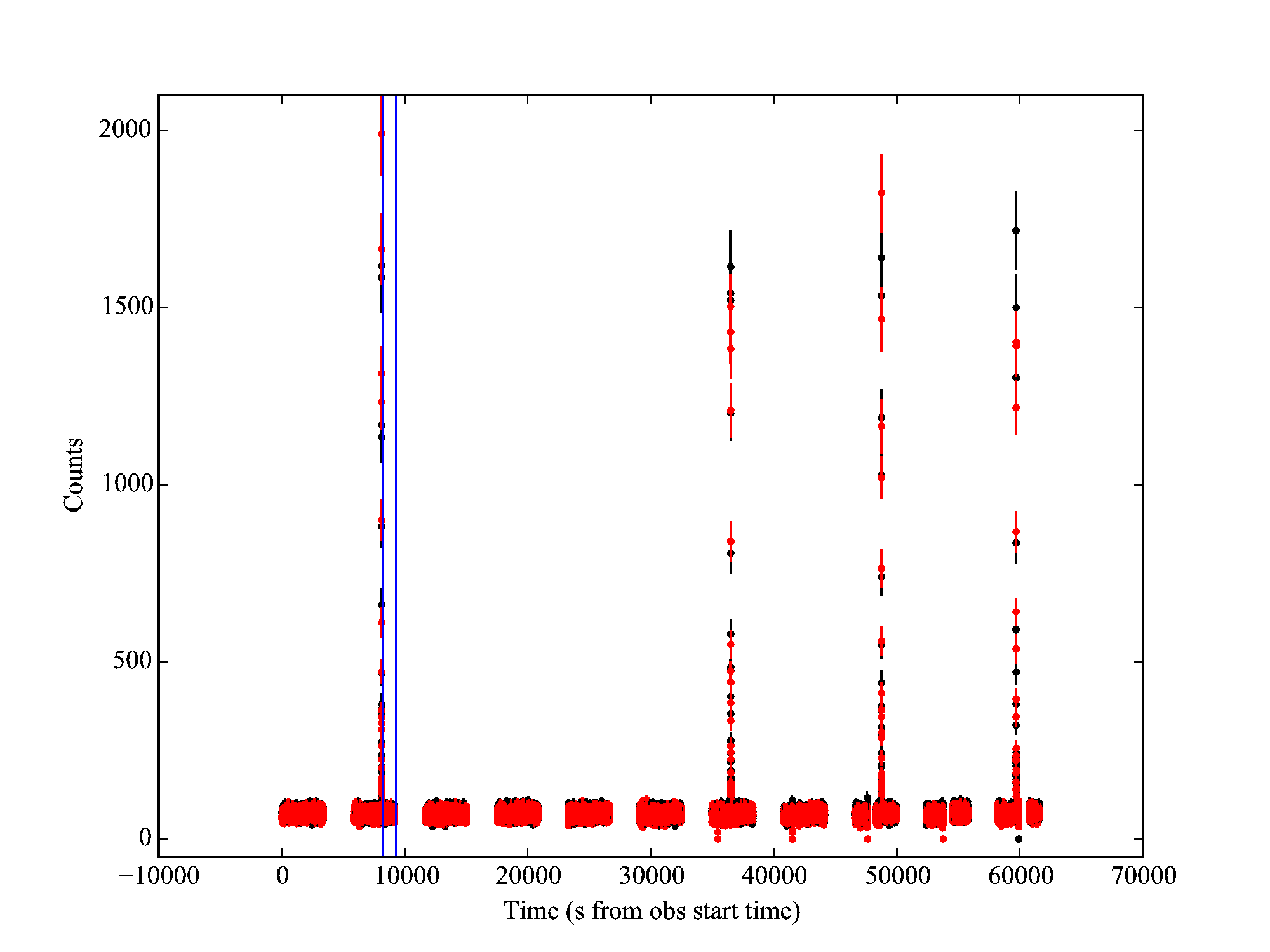}
\includegraphics*[width=0.5\textwidth]{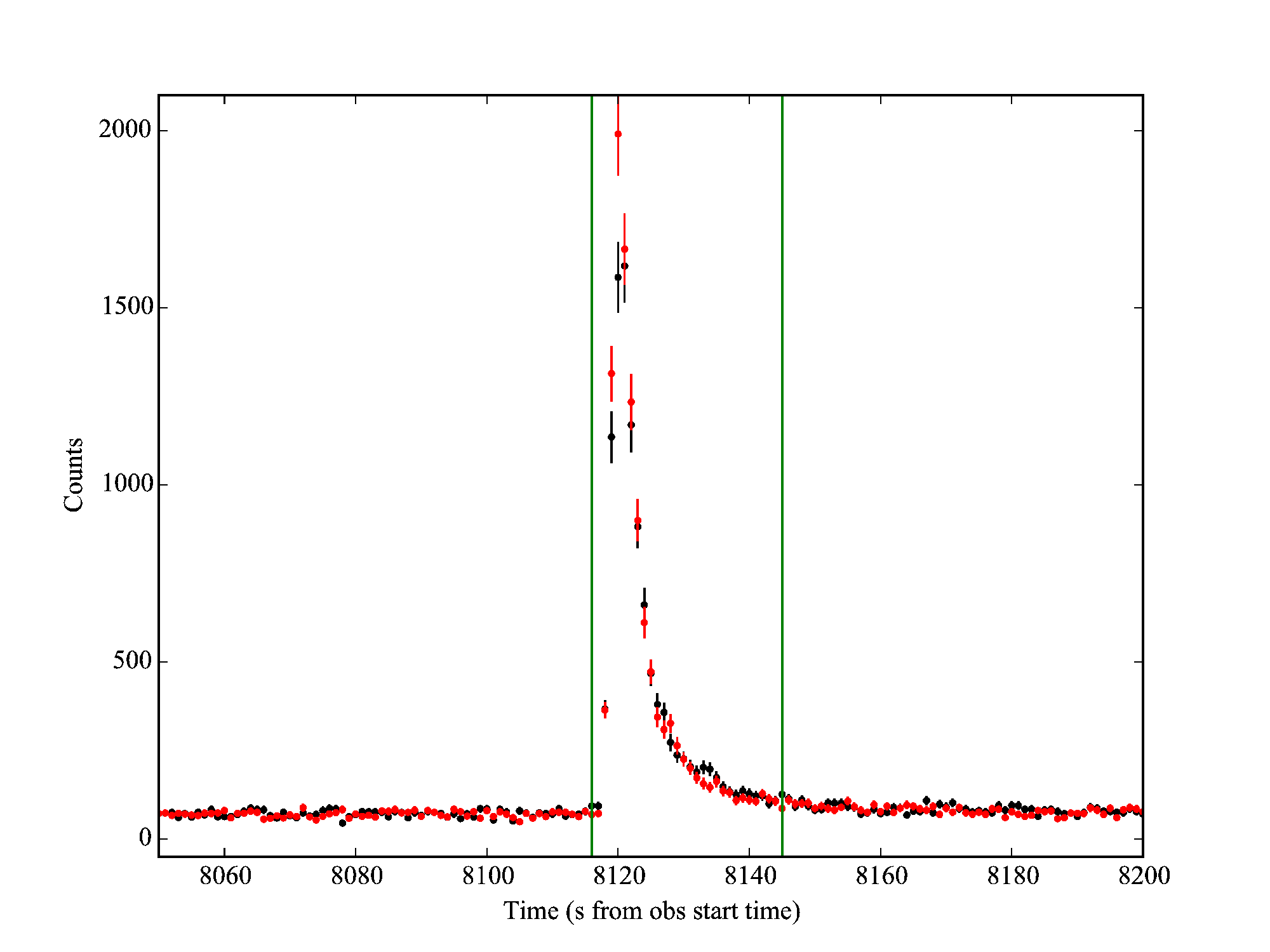}
\caption{\emph{(a)} The \emph{NuSTAR} FPMA (black) and FPMB (red) light curve, including the four type I X-ray bursts. The time of the \emph{Swift} observation is marked in blue.  \emph{(b)} A close-up of the first type I X-ray burst. The green lines indicate the data removed for spectral analysis. \label{fig:lightcurve}}
\end{figure}

\begin{figure}
\includegraphics*[width=0.5\textwidth]{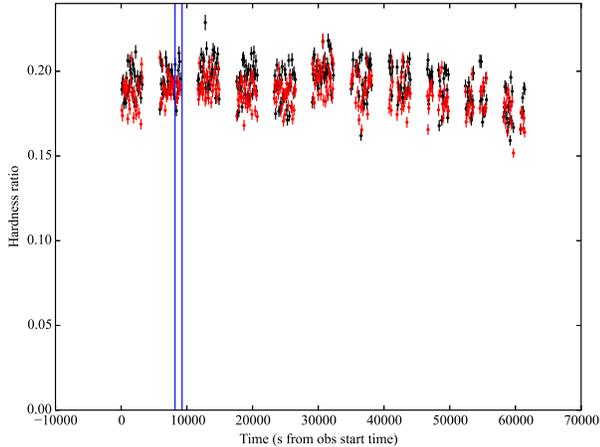}
\caption{The hardness ratio throughout the \emph{NuSTAR} observation, with \emph{NuSTAR} FPMA in red and \emph{NuSTAR} FPMB in black. The time of the \emph{Swift} observation is marked in blue. The hardness ratio is defined as the count rate from 12-25 keV / the count rate from 4.5-12 keV. \label{fig:hardness}}
\end{figure}

4U 1728-34 is known to exhibit Type-1 X-ray bursts. Using the light curves made by the {\ttfamily nuproducts FTOOL}, we detected and removed four bursts in the \emph{NuSTAR} data, each lasting about 20 s (Figure \ref{fig:lightcurve}). No bursts were detected during the \emph{Swift} observation. \cite{mondal16} did a full analysis of the bursts. To check the stability of the energy spectrum during the observation, we looked at the hardness ratio, defined here as  the count rate from 12-25 keV divided by the count rate from 4.5-12 keV (Figure \ref{fig:hardness}). The hardness ratio only changes slightly, softening gradually with time, indicating a fairly stable spectrum. By studying the power spectrum of this observation, \cite{mondal16} identified the state as the island state, and we do not disagree with this result.

\begin{figure}
\includegraphics[width=0.5\textwidth]{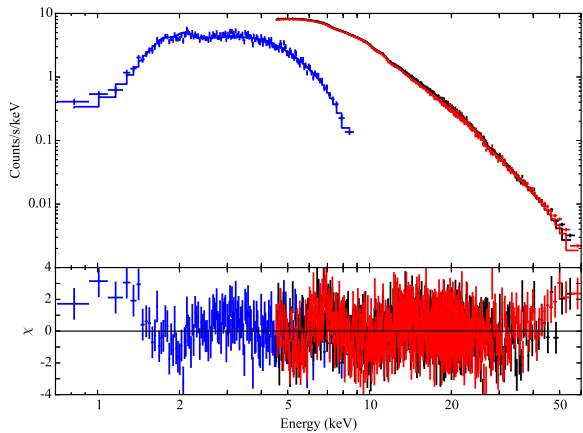}
\caption{The 4.5-78.4 keV \emph{NuSTAR} FPMA (black) and \emph{NuSTAR} FPMB (red) spectra and the 0.7-10 keV \emph{Swift} (blue) spectrum of 4U 1728-34, fit with the continuum model: disk blackbody + blackbody + power-law. The residuals between 6-8 keV indicate the presence of iron emission.\label{fig:continuum}}
\end{figure}

We first fit the combined \emph{Swift} and \emph{NuSTAR} continuum spectra similarly to \cite{ng10}, with a model consisting of a neutral absorption component {\ttfamily tbabs}, a single temperature blackbody {\ttfamily bbodyrad}, a disk blackbody {\ttfamily diskbb}, and a powerlaw component {\ttfamily cutoffpl}. This model takes into account thermal emission from the boundary layer between the neutron star surface and disk, thermal emission from the disk, and non-thermal emission from Comptonization. We found a blackbody temperature of $1.41\pm0.01$ keV, a disk blackbody temperature of $0.32\pm0.02$ keV, a photon index of $1.29\pm0.05$, and a cutoff energy of $18.3\pm0.7$ keV (all errors are 90\% confidence). This model fits most of the energy band fairly well ($\chi^2/\rm dof=1700/1257=1.35$), but there are significant residuals between 6-8 keV (Figure \ref{fig:continuum}).

We added a Gaussian line to model the excess at $\sim6$ keV. Leaving the line parameters free gives unphysical results (the line energy is at $\simeq 5.7$ keV with a width $\sigma\simeq1.6$ keV). Similarly to \cite{dai06} and \cite{egron11}, we limited the line width to $\sigma=0.5$ keV, resulting in a line energy of $6.56\pm0.05$ keV and an equivalent width of 74 eV. The addition of a Gaussian significantly improved the fit, with a $\chi^2/\rm dof = 1529/1253 = 1.22$. Moreover, the {\ttfamily diskbb} component became statistically insignificant, its exclusion from the fit resulting in $\Delta\chi^2 = 5$ and only slight changes to the other parameters. We continued our analysis without the {\ttfamily diskbb} component.

Having confirmed the presence of the iron line with the Gaussian model, we replaced the Gaussian with a physical reflection model, {\ttfamily reflionx} \citep{reflionx}.  The {\ttfamily reflionx} model used here is a version of \cite{reflionx} that assumes reflection of a power law with a high-energy exponential cutoff. To take relativistic blurring into account, we convolved {\ttfamily reflionx} with {\ttfamily kerrconv} \citep{kerrconv}.

The {\ttfamily reflionx} parameters include the photon index and cutoff energy (tied to those of the {\ttfamily cutoffpl} component), the ionization ($\xi=L/nr^2$, or the ratio of the flux to the column density, where $L$ is the luminosity, $r$ is the distance, and $n$ is the column density), the iron abundance ($A_{\rm Fe}$), and the flux. The {\ttfamily kerrconv} parameters include the compact object spin parameter, disk inclination, disk inner and outer radius, and inner and outer emissivity indices. The spin parameter $a\equiv cJ/GM^2=0.15$ (where $J$ is angular momentum) can be calculated from previous measurements of the neutron star spin frequency, 363 Hz \citep{strohmayer96}, assuming a typical neutron star mass of $M=1.4\Msun$. We fixed the disk outer radius $R_{\rm out}=400 R_{\rm ISCO}$ (where $R_{\rm ISCO}$ is the radius of the innermost stable circular orbit) because the emissivity profile drops steeply with radius, causing the fits to be insensitive to the exact value of this parameter.


As it was difficult to constrain the emissivity of the reflecting disk when it was left free, we considered fits with $q=1,2,3,4$ and $5$. We also considered modeling the emissivity as a broken power law with the outer index fixed to 3, the inner index left free, and the break radius fixed to $25\rm R_{\rm g}$, but we were unable to constrain the inner index. As shown in Table \ref{tab:emissivity}, the changes in emissivity negligibly affect the $\chi^2$. Furthermore, the values for the inner radius are very similar across the models (other than $q=1$, where the inner radius has very large error bars). As the parameters, particularly the inner radius, seem to be insensitive to the emissivity, we continued our analysis with the emissivity fixed at $q=3$, consistent with a Newtonian geometry far from the neutron star.

\begin{figure*}
\includegraphics*[width=0.5\textwidth]{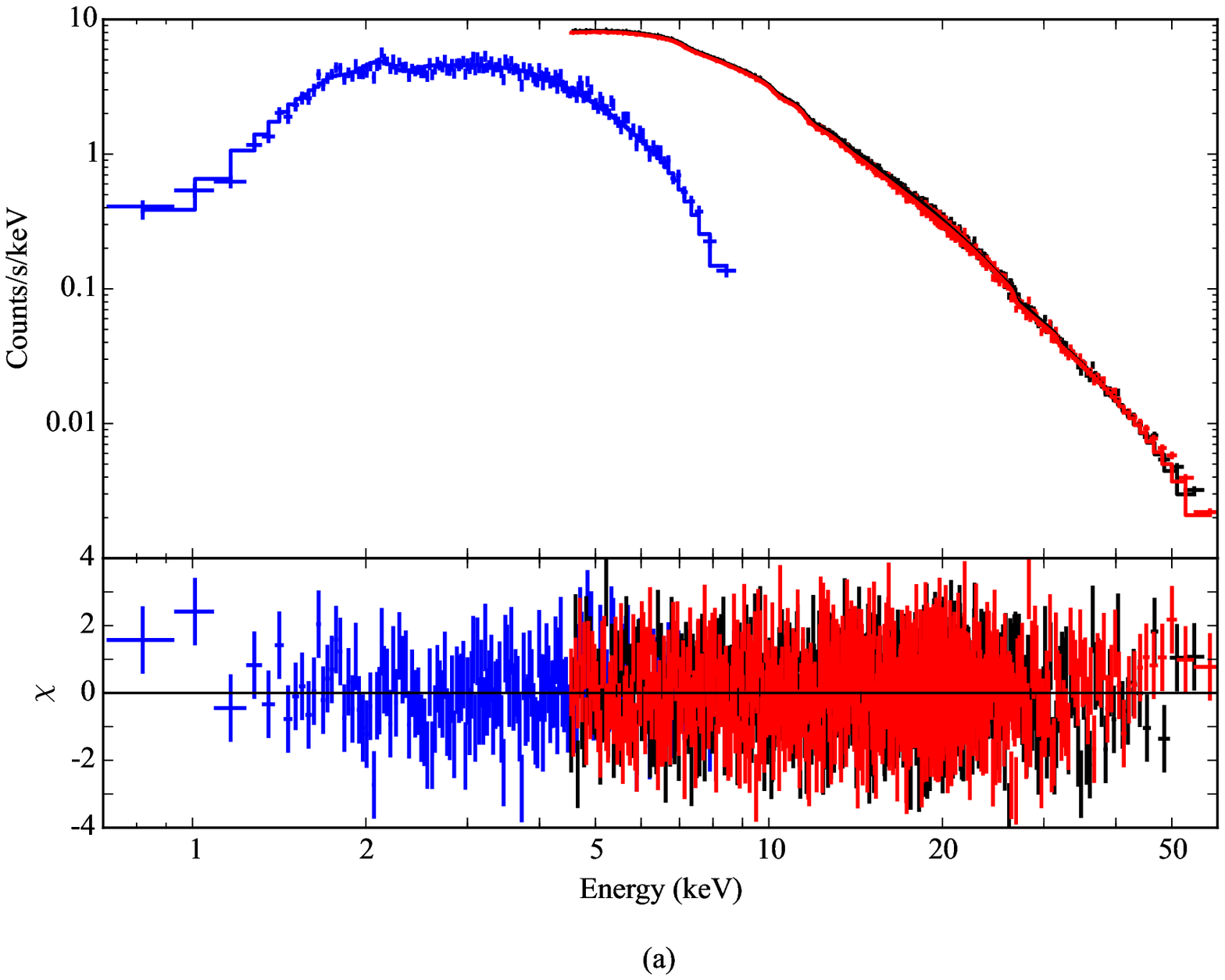}
\includegraphics*[width=0.5\textwidth]{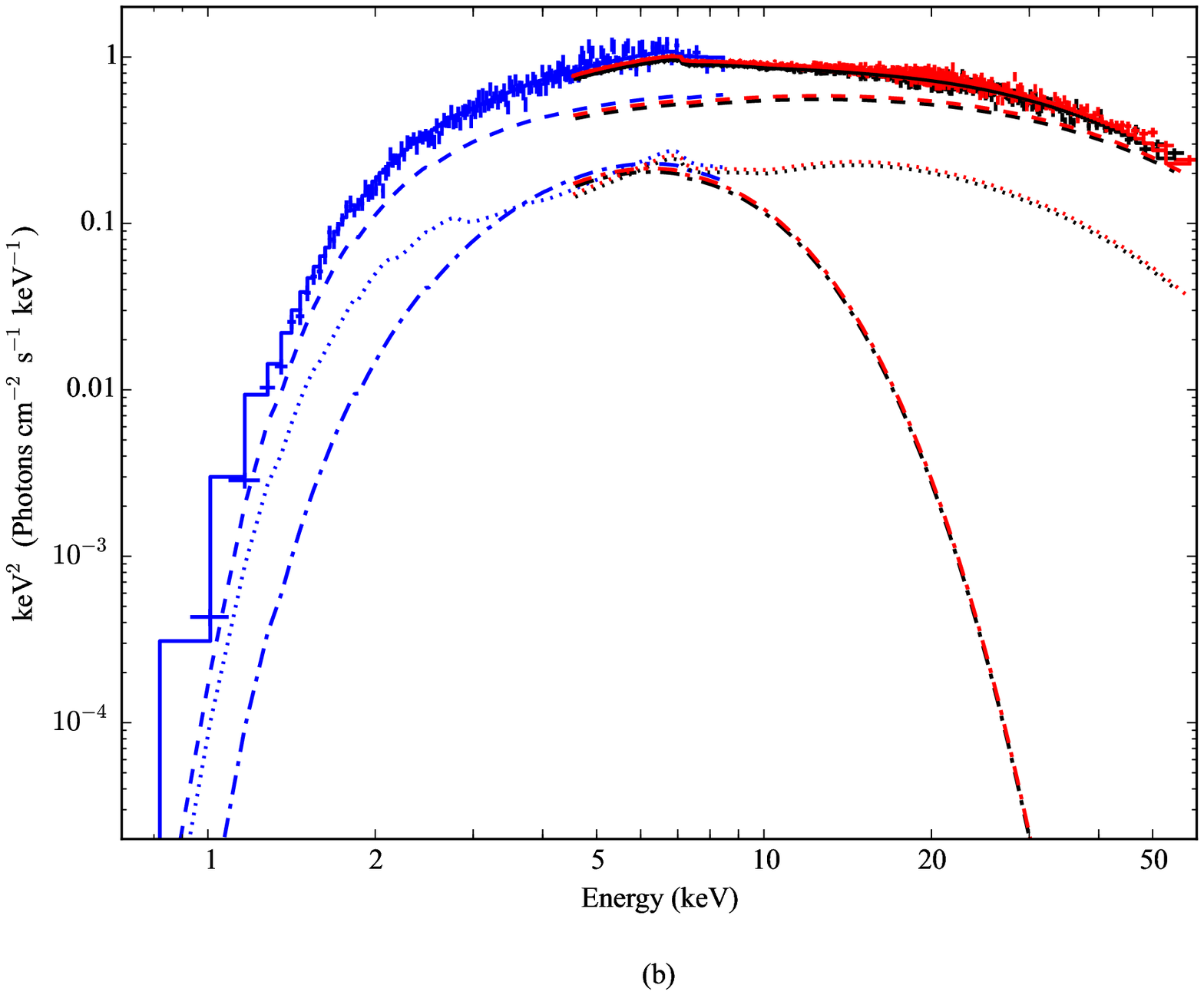}
\caption{The 4.5-78.4 keV \emph{NuSTAR} FPMA (black) and \emph{NuSTAR} FPMB (red) spectra and the 0.7-10 keV \emph{Swift} (blue) spectrum of 4U 1728-34, fit with the {\ttfamily reflionx} relativistically blurred reflection model (model 1). 
\emph{(a)} Shows the residuals of the reflection model, specifically the lack thereof between 6-8 keV. \emph{(b)} Shows the $\nu F_\nu$ plot with individual model components: blackbody (dashed and dotted), cutoff power-law (dashed), and {\ttfamily reflionx} (dotted).\label{fig:reflionx}}
\end{figure*}

\begin{figure*}
\includegraphics*[width=0.5\textwidth]{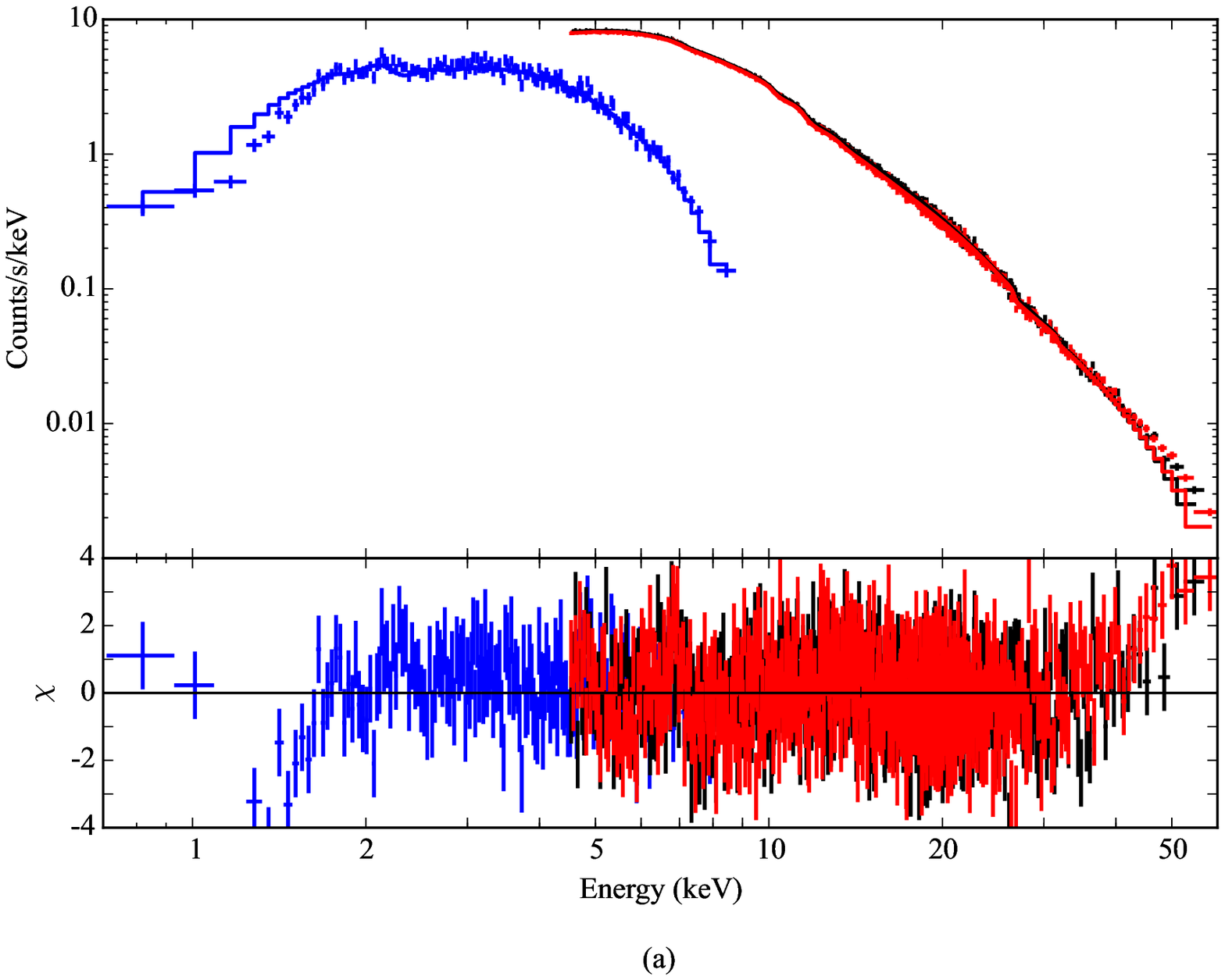}
\includegraphics*[width=0.5\textwidth]{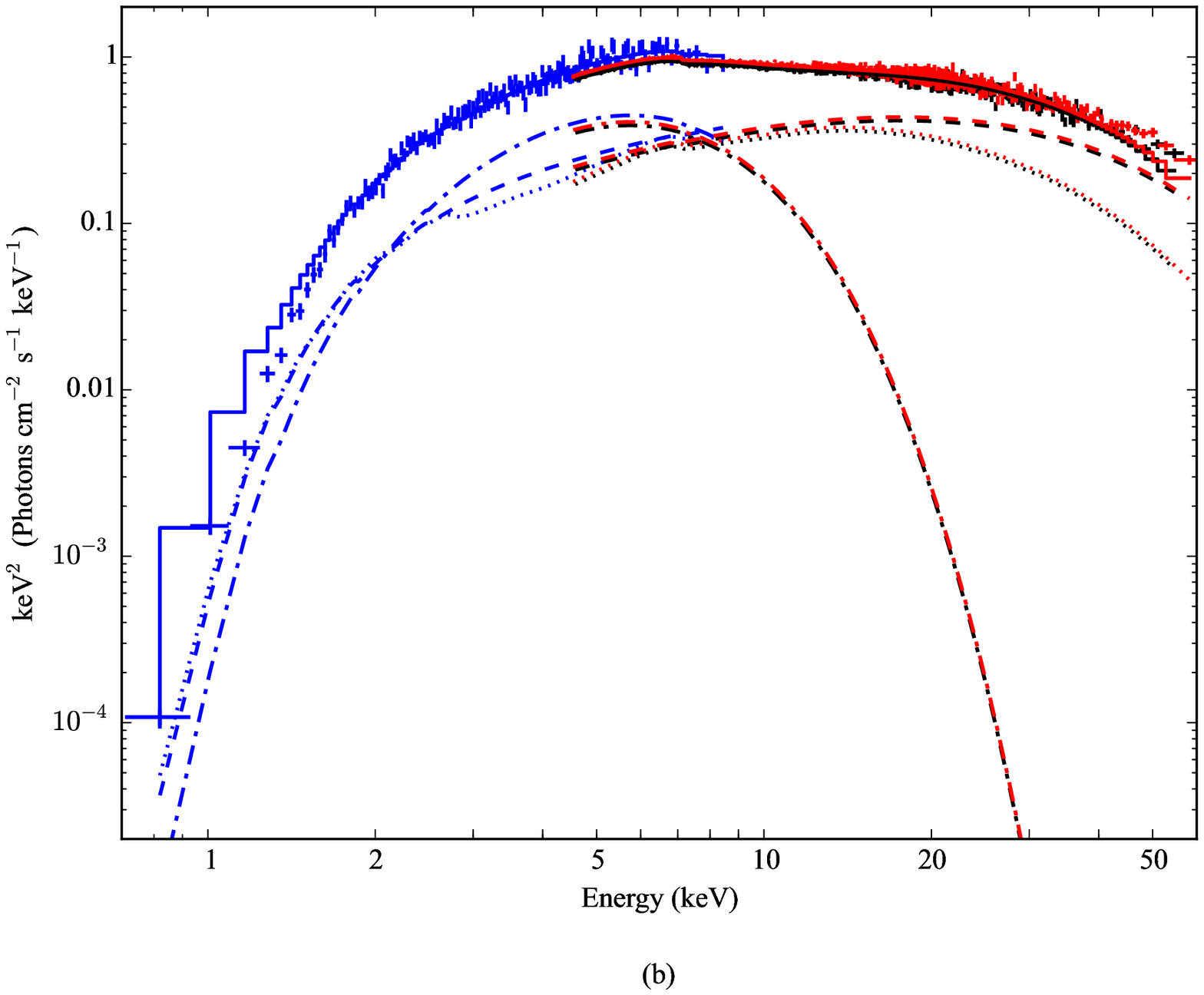}
\caption{The 4.5-78.4 keV \emph{NuSTAR} FPMA (black) and \emph{NuSTAR} FPMB (red) spectra and the 0.7-10 keV \emph{Swift} (blue) spectrum of 4U 1728-34, fit with the {\ttfamily reflionx} relativistically blurred reflection model where the column density is fixed to $2.9\times10^{22}$ cm$^{-2}$ (model 2). 
\emph{(a)} Shows the residuals of the reflection model. \emph{(b)} Shows the $\nu F_\nu$ plot with individual model components: blackbody (dashed and dotted), cutoff power-law (dashed), and {\ttfamily reflionx} (dotted).\label{fig:nH_frozen}}
\end{figure*}

Adding the reflection component (Figure \ref{fig:reflionx}) improves the fit over the Gaussian line model ($\chi^2/\rm dof=1430/1254=1.14; \Delta\chi^2=99$). Table \ref{tab:parameters} lists the best fit parameters for the relativistically blurred reflection model (model 1). We find an inclination of $\sim37^{\circ}$, consistent with the lack of dips in the light curve which implies a low viewing angle. From the normalization of {\ttfamily bbodyrad}, we find a blackbody source radius of 1.4 km, consistent with thermal emission from the boundary layer. We find a higher column density, $N_{\rm H}\sim4.5\times10^{22}$ cm$^{-2}$, than what has been previously measured for this source ($N_{\rm H}\sim2.6-2.9\times10^{22}$ cm$^{-2}$, e.g. \citealt{disalvo00,piraino00,narita01,egron11}). To be consistent with past measurements, we fixed the column density to $2.9\times10^{22}$ cm$^{-2}$ (Figure \ref{fig:nH_frozen}; model 2 in Table \ref{tab:parameters}). This significantly worsens the fit, resulting in a $\chi^2/\rm dof=1712/1255=1.36$ ($\Delta\chi^2=282$), yet does not cause a large change to the main parameter of interest, $R_{\rm in}$.

Due to the proximity of 4U 1728-34 to the Galactic plane ($b=-0.15^{\circ}$), it is reasonable to consider the column density of molecular hydrogen. Galactic surveys indicate that at the position of 4U 1728-34, $N_{\rm H_2}\sim1.8\times10^{22}$ cm$^{-2}$ \citep{dame01} and $N_{\rm H_I}\sim1.24\times10^{22}$ cm$^{-2}$ \citep{kalberla05}. Thus, the expected total column density is $N_{\rm H_{\rm total}} = N_{\rm H_I} + 2N_{\rm H_2} \sim 4.84 \times 10^{22}$ cm$^{-2}$, close to our measured value. We conclude that the model with the column density as a free parameter, which is a better fit to the data, is more correct. From this model, we find an upper limit for the disk inner radius $R_{\rm in}\leq1.77 R_{\rm ISCO}$, with the best value at $R_{\rm in} = 1.00^{+0.77}_{-0} R_{\rm ISCO}$.

We consider if the reflection parameters are mostly constrained by the iron line as opposed to the reflected continuum by fitting the data with the {\ttfamily relline} model, a relativistic line profile excluding broadband features such as the Compton hump. The best fit parameters are shown in model 3 of Table \ref{tab:parameters}. The value for the inner radius, $R_{\rm in}=1.1^{+0.2}_{-0}R_{\rm ISCO}$, is consistent with our above upper limit of $R_{\rm in}\leq1.77R_{\rm ISCO}$ and is even better constrained. However, the {\ttfamily relline} model, with $\chi^2/\rm dof=1532/1255=1.22$, is not as good a fit as the broadband reflection model described above ($\chi^2/\rm dof=1.14$), suggesting that the broadband reflection spectrum does make some contribution, at least statistically.

To verify that the \emph{Swift} spectrum helps constrain the spectral shape, we fit only the \emph{NuSTAR} data using the {\ttfamily reflionx} model (model 4 in Table \ref{tab:parameters}). We fixed the column density to $2.9\times10^{22}$ cm$^{-2}$ as the \emph{NuSTAR} data is unable to measure this parameter. Fitting only the \emph{NuSTAR} spectrum with this model resulted in a $\chi^2/\rm dof=1256/1096=1.15$ and is statistically better than including the \emph{Swift} data 
(model 3 in Table \ref{tab:parameters}; $\chi^2/\rm dof=1712/1255=1.36$). However, without the low-energy coverage offered by \emph{Swift}, the column density cannot be well measured and thus the inaccuracy of fixing $N_{\rm H}=2.9\times10^{22}$ cm$^{2}$ is not revealed in the fit statistics. Without the \emph{Swift} spectrum, we find $R_{\rm in}\leq2.7R_{\rm ISCO}$ as an upper limit for the inner radius (contrasting $R_{\rm in}\leq1.4R_{\rm ISCO}$ found in model 3), indicating that the \emph{Swift} spectrum is useful in evaluating the main parameter of interest as well as the column density.

\cite{mondal16} have also analyzed this coordinated \emph{NuSTAR} and \emph{Swift} observation, fitting the data instead with {\ttfamily relxill}, another relativistically blurred reflection model \citep{relxill}. To compare to their results, we also fit the spectrum with {\ttfamily relxill}, and in addition replaced {\ttfamily cutoffpl} for the more physical {\ttfamily comptt}. The {\ttfamily relxill} parameters include the ionization, iron abundance, compact object spin parameter, disk inclination, disk inner and outer radius, and inner and outer emissivity indices. We fixed $R_{\rm out}=400R_{\rm g}$ and $a=0.15$. Similarly to \cite{mondal16}, we fixed $q_{\rm out}=3$ and left $q_{\rm in}$ free, but we were only able to find a lower limit on $q_{\rm in}$. The {\ttfamily relxill} model is statistically similar to model 1 with a $\chi^2/\rm dof=1412/1250=1.13$, and the best fit parameters shown in Table \ref{tab:relxill} are comparable to those of model 1. In particular, the column density $N_{\rm H}=3.9\times10^{22}$cm$^{-2}$ is considerably higher than past measurements and the inclination $i=37^{\circ}$ is the same as model 1. We find an upper limit of $R_{\rm in}\leq2R_{\rm ISCO}$ for the disk inner radius, with the best value at $R_{\rm in}=1.6\pm0.4$. This upper limit is close to our previous upper limit of $1.77R_{\rm ISCO}$, but as it is slightly higher, we continue our analysis considering $R_{\rm in}\leq2R_{\rm ISCO}$ as the upper limit for the disk inner radius.

\begin{figure*}
\includegraphics*[width=0.5\textwidth]{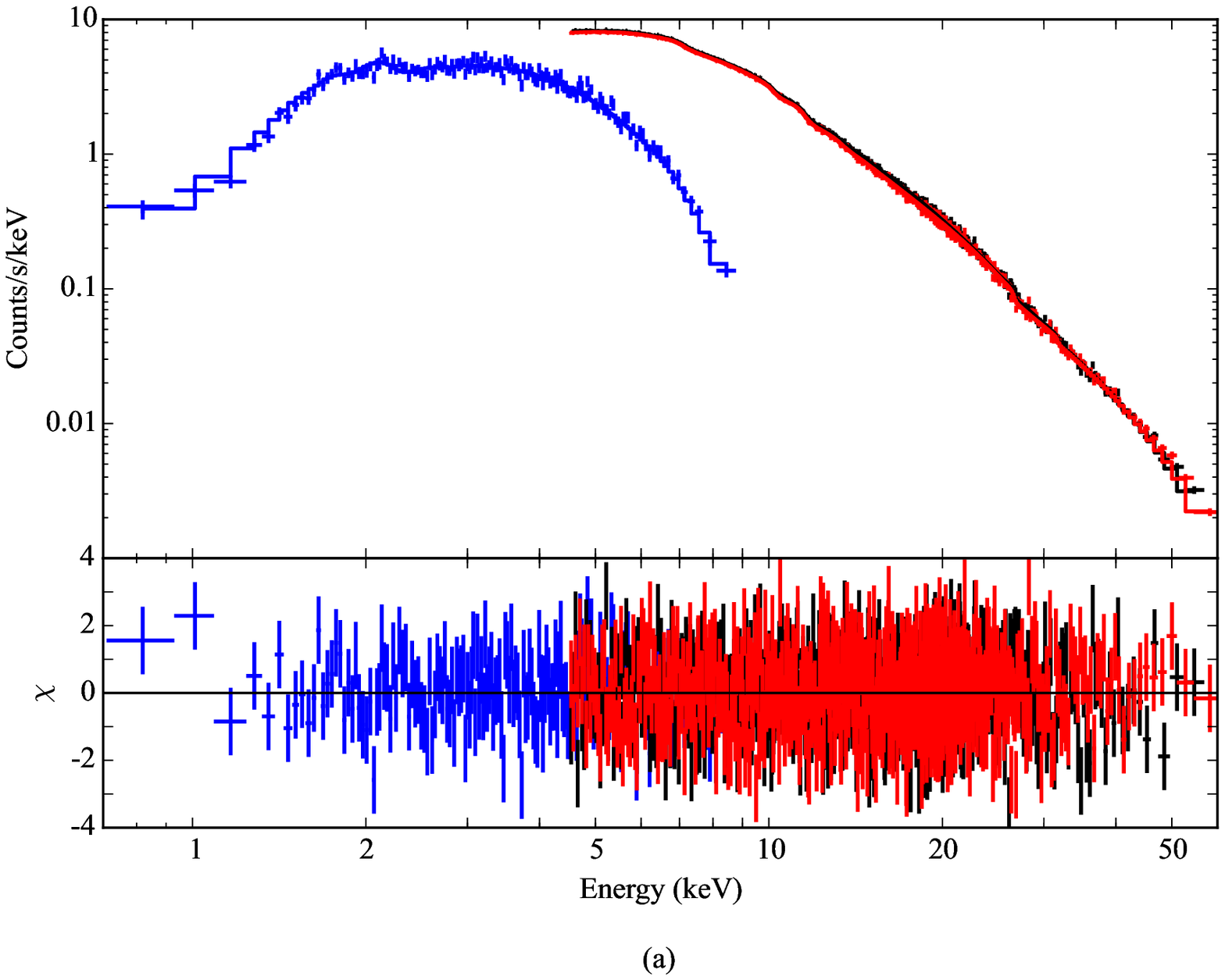}
\includegraphics*[width=0.5\textwidth]{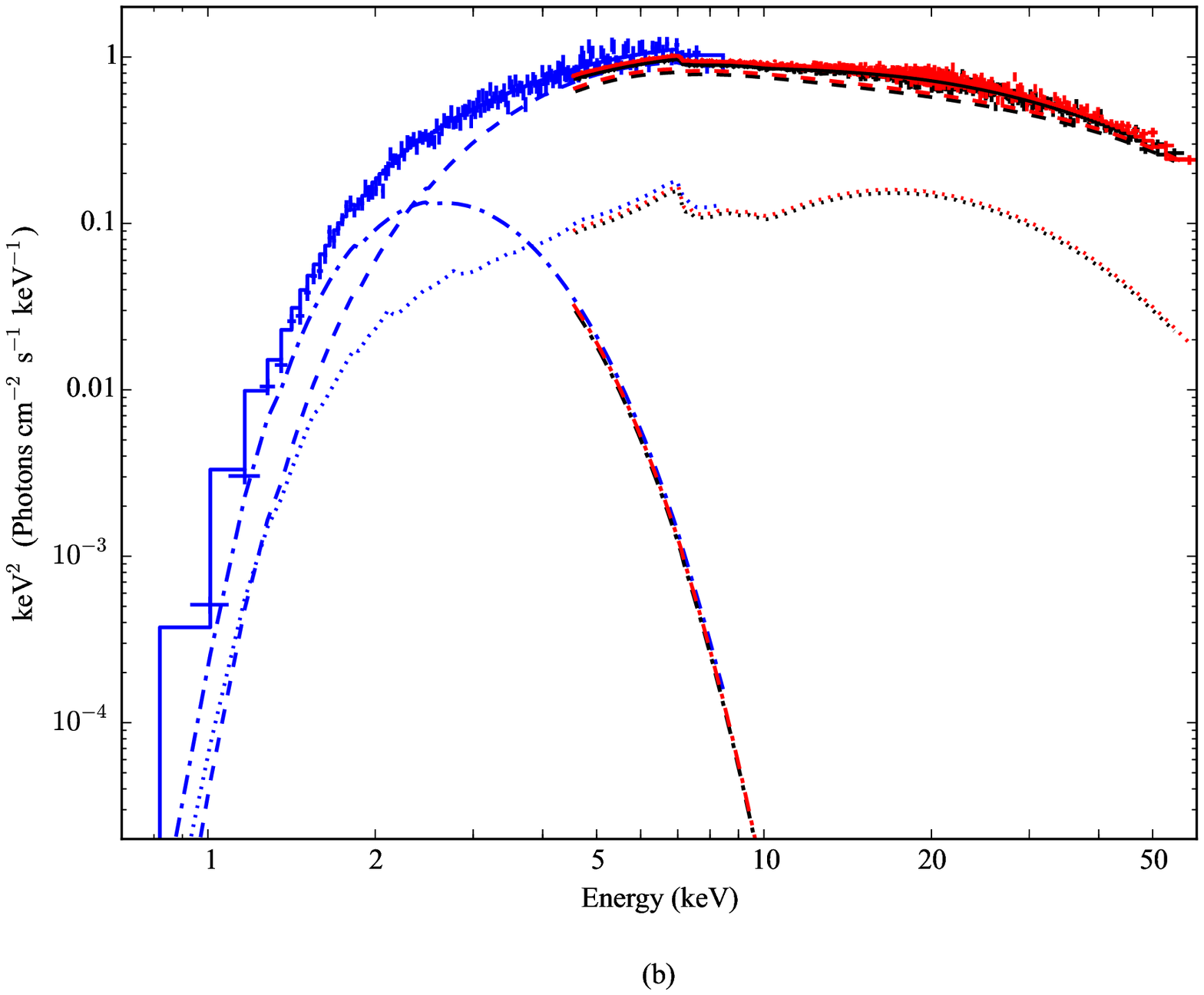}
\caption{The 4.5-78.4 keV \emph{NuSTAR} FPMA (black) and \emph{NuSTAR} FPMB (red) spectra and the 0.7-10 keV \emph{Swift} (blue) spectrum of 4U 1728-34, fit with {\ttfamily relxill} (model 5). (a) Shows the residuals of the reflection model. (b) Shows the $\nu F_\nu$ plot with individual model components: blackbody (dashed and dotted), {\ttfamily comptt} (dashed), and {\ttfamily relxill} (dotted).\label{fig:relxill}}
\end{figure*}

\begin{figure*}
\includegraphics*[width=0.5\textwidth]{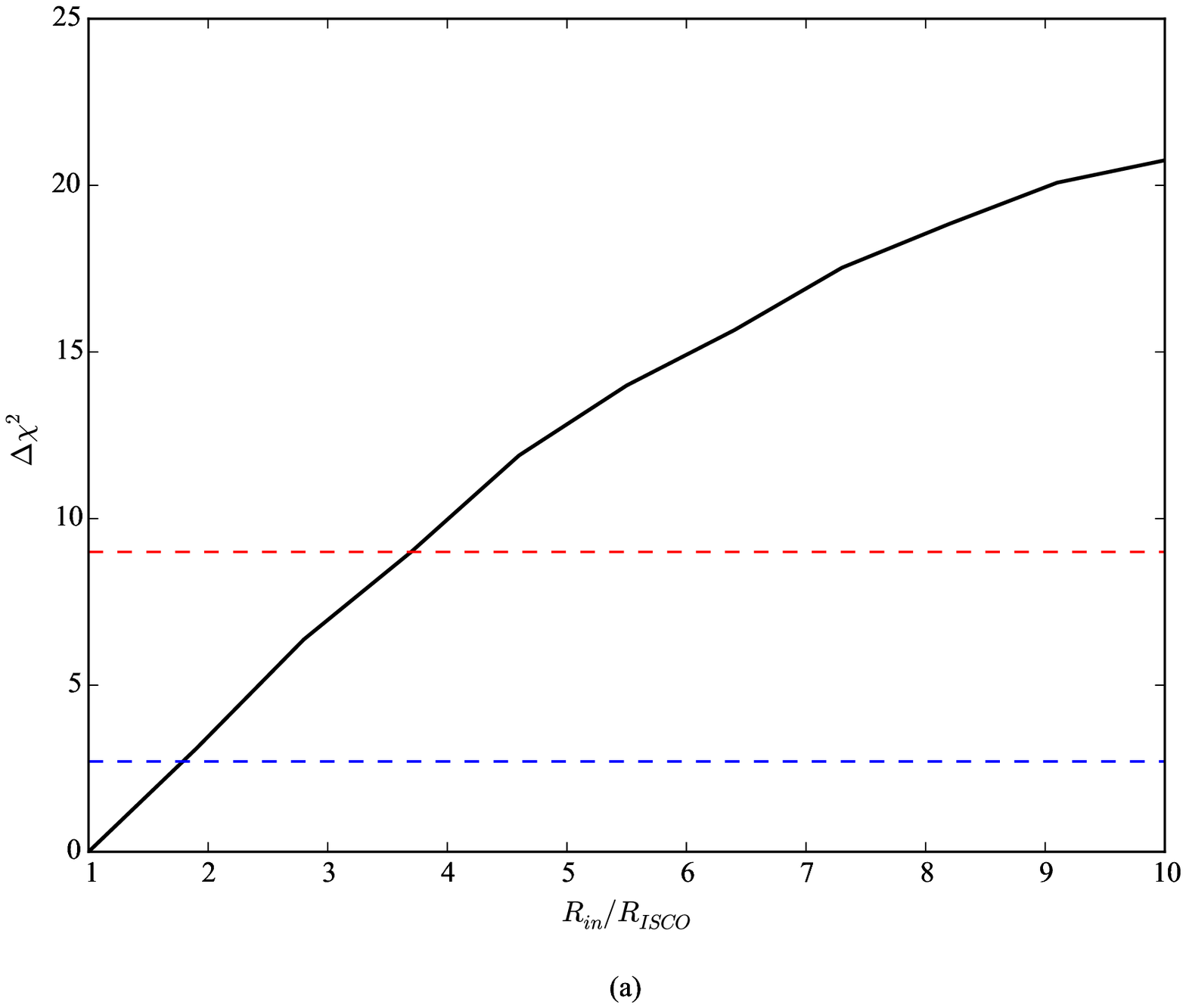}
\includegraphics*[width=0.5\textwidth]{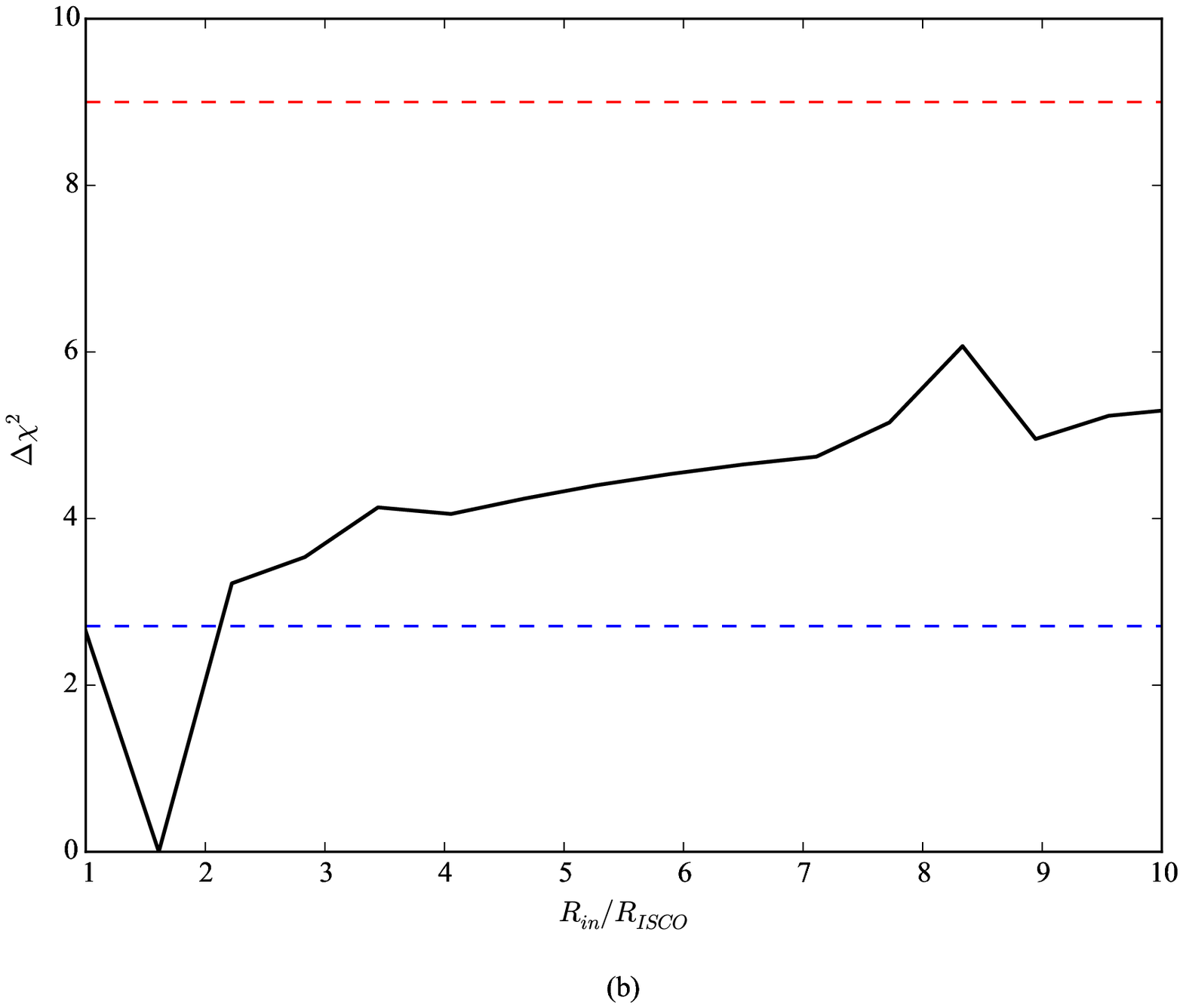}
\caption{The change in $\chi^2$ versus the change in inner radius, showing that the model is consistent with the disk extending to the ISCO. The dashed blue line marks 90\% confidence and the dashed red line marks 99.7\% confidence ($3\sigma$). \emph{(a)} Shows the results of the {\ttfamily reflionx} model (model 1). \emph{(b)} Shows the results of the {\ttfamily relxill} model (model 5).
\label{fig:rin}}
\end{figure*}

With spin parameter value of $a=0.15$, we calculate 1 $R_{\rm ISCO} = 5.5R_g$ \citep{bardeen72}. Assuming a typical neutron star mass of $1.4\Msun$, we find an upper limit for the neutron star radius $R_{NS}\leq2\times5.5R_g=23$ km (Figure \ref{fig:rin}).

\section{Discussion}

The unabsorbed flux extrapolated in the 0.1-100 keV energy range is $F\sim6\times10^{-9}$ erg cm$^{-2}$ s$^{-1}$. Assuming a distance to the source of 5.1 kpc \citep{disalvo00,galloway03}, we calculate a luminosity $L_{0.1-100} = 1.9\times10^{37}$ erg s$^{-1}$, corresponding to 8\% of the Eddington luminosity, $L_{\rm EDD} = 2.5\times10^{38}$ erg s$^{-1}$ \citep{vanparadijs94}. We note that this value is around an average luminosity for 4U 1728-34 compared to previous observations. \cite{disalvo00} measured $L_{0.1-100}=2\times10^{37}$ erg s$^{-1}$. Others have used different energy ranges; for the sake of comparison, we re-calculate the unabsorbed flux and luminosity in different ranges. \cite{egron11} measured $L_{0.1-150}=5\times10^{36}$ erg s$^{-1}$, in comparison to our value of $L_{0.1-150}=1.9\times10^{37}$ erg s$^{-1}$. \cite{piraino00} measured $L_{0.2-50}=3.7\times10^{37}$ erg s$^{-1}$, while we measure of $L_{0.2-50}\sim1.7\times10^{37}$ erg s$^{-1}$. \cite{ng10} measured $L_{2-10}=7.8\times10^{35}$ erg s$^{-1}$, compared to our measurement of $L_{2-10}\sim6.6\times10^{36}$ erg s$^{-1}$.

With our upper limit on the accretion disk inner radius $R_{\rm in}\leq23$ km, we infer an upper limit on the neutron star magnetic field strength by equating the magnetic field pressure to the ram pressure of the accretion disk \citep{bfield}. We use the following relation:
\begin{equation}
R_{\rm in} = 4\times10^8B_{11}^{4/7}\dot{m}_{15}^{-2/7}M^{-1/7}\textnormal{ cm}
\end{equation}
where $B_{11}$ is the magnetic field in units of $10^{11}$ G, $\dot{m}_{15}$ is the mass accretion rate in units of $10^{15}$ g s$^{-1}$, and $M$ is the neutron star mass in solar masses. We calculate the mass accretion rate from the luminosity given above with the relation $L=\eta\dot{m}c^2$, where $\eta=GM/Rc^2$. Assuming $M=1.4\Msun$, we find $B\leq2\times10^8$ G.

We compare our results with those obtained for other neutron star LMXBs also observed with \emph{NuSTAR}. \cite{miller13} found that Serpens X-1, a persistent source, has a disk extending almost to the ISCO when observed at a luminosity of $L\sim0.44L_{\rm EDD}$. From their measured values of $L$ and $R_{\rm in}$, we estimate $B\leq2.2\times10^8$ G. \cite{degenaar15} found that 4U 1608-52, a transient source with a spin frequency of 620 Hz, also has a disk extending close to the ISCO when observed at a luminosity of $L\sim0.02L_{\rm EDD}$; we estimate $B\leq1.2\times10^8$ G. Aql X-1, also a transient observed by \cite{king15} at a luminosity of $L\sim0.08L_{\rm EDD}$, has a truncated disk ($R_{\rm in}\sim15R_g$), a spin frequency of 550 Hz, and a magnetic field $B\leq5\pm2\times10^8$ G.

In using Equation 1, we assume that the magnetic pressure truncates the accretion disk \citep{bfield}. If this is the case, we expect to see pulsations, yet of the above sources only Aql X-1 has been observed to exhibit pulsations during outburst \citep{casella08}. \cite{papitto13} note that exhibiting pulsations is rare among LMXBs and consider various explanations, including the possibility that the spin axis is aligned with the magnetic axis \citep{lamb09}, that pulsations do exist but the pulse amplitude is below the detectable threshold, or that magneto-hydrodynamical instabilities cause material to accrete onto the neutron star at random places instead of being channeled by the magnetic field lines \citep{romanova08}. The lack of pulsations in 4U 1728-34, 4U 1608-52 and Serpens X-1 may indicate that our assumption of the magnetic pressure truncating the disk is not correct or at least not the complete physical picture. In that case, the true magnetic field values of these sources are most likely somewhat lower than the upper limits quoted above.

According to \cite{white97}, a higher spin frequency should imply a lower magnetic field: with a low magnetic field, the disk can extend deeper into the potential, spinning up the neutron star. While it is consistent that the disk extends closer to the ISCO in the above sources with lower magnetic fields, the correlation between spin frequency and magnetic field is not always followed. While 4U 1608-52 has the highest spin frequency and lowest magnetic field, Aql X-1 has both a higher spin frequency and a higher magnetic field than 4U 1728-34. (The spin frequency of Serpens X-1 is unknown, so we leave it out of this comparison.)

It is likely that this discrepancy is due to the expected variability in LMXB spin periods and magnetic fields. Additionally, as we are considering upper limits on the magnetic fields, we recognize that better estimates of the magnetic field values could negate this discrepancy. It is possible, however, that the discrepancy is real, in which case we consider the effects of magnetic field screening \citep{cumming01} as an explanation. Magnetic field screening is a process by which the accreting matter becomes magnetized slowly compared to the accretion rate, causing the field implied by equating the pressures to be orders of magnitude smaller than the true field of the neutron star. Because Aql X-1 is a transient, we consider the possibility that the magnetic field emerges during quiescence and is not screened immediately after the outburst begins. The time it takes for the magnetic field to become significantly screened depends on accretion rate; for an accretion rate of $0.08L_{\rm EDD}$, the field will be screened by one order of magnitude five days or less after the outburst began \citep{cumming08}. The observation of Aql X-1 occurred less than five days after the outburst began; hence it is possible that Aql X-1 is not fully screened (though a better estimate of the screening timescale could further support, or refute, this hypothesis). This possibility provides an explanation for the low spin frequency and magnetic field of 4U 1728-34 as compared to Aql X-1.


We consider possible mechanisms for accretion onto the neutron star. Because the magnetic field is relatively low, it is possible that the material is getting to the neutron star via a magnetic gate \citep{lamb77}. Due to chaotic accretion on the stellar surface, type II X-ray bursts are expected in the magnetic gate model, yet are not exhibited in 4U 1728-34. Instead, the material could be channeled along the magnetic field lines to the poles \citep{lamb09}. This scenario would cause a hot spot on the magnetic pole. However, it is possible that the magnetic axis is aligned with the rotation axis as 4U 1728-34 does not emit regular pulsations.

\section{Summary}
We have analyzed the persistent emission of neutron star LMXB 4U 1728-34 using a concurrent $NuSTAR$ and $Swift$ observation. By fitting the continuum with a power law and thermal components, we find clear evidence of an Fe K$\alpha$ line in the spectrum of 4U 1728-34. With a relativistically blurred reflection model, we find an upper limit to the accretion disk inner radius, and thus the neutron star radius, of $R_{NS}\leq23$ km. From this, we infer the upper limit on the magnetic field of the neutron star to be $B\leq2\times10^8$ G.

\acknowledgments
We thank Michael Parker and Andy Fabian for the particular version of the {\ttfamily reflionx} model used in this analysis. We thank Kristin Madsen for her help identifying the calibration issue in the \emph{NuSTAR} data between 3-4.5 keV. JC thanks ESA/PRODEX for financial support. This work is based on data from the \emph{NuSTAR} mission, a project led by the California Institute of Technology, managed by the Jet Propulsion Laboratory, and funded by NASA.

\bibliographystyle{jwapjbib}
\bibliography{refs}

\begin{thebibliography}{}

\bibitem[\protect\astroncite{{Arnaud}}{1996}]{xspec}
{Arnaud}, K.~A.,  1996,
\newblock in Astronomical Data Analysis Software and Systems V, ed. G.~H.
  {Jacoby}, J. {Barnes}, Vol. 101, Astronomical Society of the Pacific
  Conference Series, ~17

\bibitem[\protect\astroncite{{Bardeen} et~al.}{1972}]{bardeen72}
{Bardeen}, J.~M., {Press}, W.~H., \& {Teukolsky}, S.~A.  1972, ApJ, 178, 347

\bibitem[\protect\astroncite{{Brenneman} \& {Reynolds}}{2006}]{kerrconv}
{Brenneman}, L.~W., \& {Reynolds}, C.~S.  2006, ApJ, 652, 1028

\bibitem[\protect\astroncite{{Burrows} et~al.}{2005}]{burrows05}
{Burrows}, D.~N., {Hill}, J.~E., {Nousek}, J.~A., et~al.\  2005, Space Science
  Reviews, 120, 165

\bibitem[\protect\astroncite{{Cackett} et~al.}{2010}]{cackett10}
{Cackett}, E.~M., {Miller}, J.~M., {Ballantyne}, D.~R., et~al.\  2010, ApJ,
  720, 205

\bibitem[\protect\astroncite{{Casella} et~al.}{2008}]{casella08}
{Casella}, P., {Altamirano}, D., {Patruno}, A., {Wijnands}, R., \& {van der
  Klis}, M.  2008, ApJ, 674, L41

\bibitem[\protect\astroncite{{Cumming}}{2008}]{cumming08}
{Cumming}, A.,  2008, AIP Conference Proceedings, 1068

\bibitem[\protect\astroncite{{Cumming} et~al.}{2001}]{cumming01}
{Cumming}, A., {Zweibel}, E., \& {Bildsten}, L.  2001, ApJ, 557, 958

\bibitem[\protect\astroncite{{D'A{\'i}} et~al.}{2006}]{dai06}
{D'A{\'i}}, A., {Di Salvo}, T., {Iaria}, R., et~al.\  2006, A\&A, 448, 817

\bibitem[\protect\astroncite{{Dame} et~al.}{2001}]{dame01}
{Dame}, T.~M., {Hartmann}, D., \& {Thaddeus}, P.  2001, ApJ, 547, 792

\bibitem[\protect\astroncite{{Degenaar} et~al.}{2015}]{degenaar15}
{Degenaar}, N., {Miller}, J.~M., {Chakrabarty}, D., et~al.\  2015, MNRAS, 451,
  L85

\bibitem[\protect\astroncite{{Di Salvo} et~al.}{2000}]{disalvo00}
{Di Salvo}, T., {Iaria}, R., {Burderi}, L., \& {Robba}, A.  2000, ApJ, 542,
  1034

\bibitem[\protect\astroncite{{Egron} et~al.}{2011}]{egron11}
{Egron}, E., {Di Salvo}, T., {Burderi}, L., et~al.\  2011, A\&A, 530, 7

\bibitem[\protect\astroncite{{Fabian} et~al.}{1989}]{fabian89}
{Fabian}, A.~C., {Rees}, M.~J., {Stella}, L., \& {White}, N.~E.  1989, MNRAS,
  238, 729

\bibitem[\protect\astroncite{{Galloway} et~al.}{2003}]{galloway03}
{Galloway}, D.~K., {Psaltis}, D., {Chakrabarty}, D., \& {Muno}, M.~P.  2003,
  ApJ, 590, 999

\bibitem[\protect\astroncite{{Galloway} et~al.}{2010}]{galloway10}
{Galloway}, D.~K., {Yao}, Y., {Marshall}, H., {Misanovic}, Z., \& {Weinberg},
  N.  2010, ApJ, 724, 417

\bibitem[\protect\astroncite{{Garc{\'i}a} et~al.}{2014}]{relxill}
{Garc{\'i}a}, J., {Dauser}, T., {Lohfink}, A., et~al.\  2014, ApJ, 782, 76

\bibitem[\protect\astroncite{{Gehrels} et~al.}{2004}]{gehrels04}
{Gehrels}, N., {Chincarini}, G., {Giommi}, P., et~al.\  2004, ApJ, 611, 1005

\bibitem[\protect\astroncite{{Gottwald} et~al.}{1995}]{gottwald95}
{Gottwald}, M., {Parmar}, A.~N., {Reynolds}, A.~P., et~al.\  1995, A\&AS, 109,
  9

\bibitem[\protect\astroncite{Harrison et~al.}{2013}]{harrison13}
Harrison, F.~A., Craig, W.~W., Christensen, F.~E., et~al.\  2013, ApJ, 770, 103

\bibitem[\protect\astroncite{{Hasinger} \& {van der Klis}}{1989}]{hasinger89}
{Hasinger}, G., \& {van der Klis}, M.  1989, A\&A, 225, 79

\bibitem[\protect\astroncite{{Illarionov} \& {Sunyaev}}{1975}]{bfield}
{Illarionov}, A.~F., \& {Sunyaev}, R.~A.  1975, A\&A, 39, 185

\bibitem[\protect\astroncite{{Kalberla} et~al.}{2005}]{kalberla05}
{Kalberla}, P.~M.~W., {Burton}, W.~B., {Hartmann}, D., et~al.\  2005, A\&A,
  440, 775

\bibitem[\protect\astroncite{{King} et~al.}{2016}]{king15}
{King}, A.~L., {Tomsick}, J.~A., {Miller}, J.~M., et~al.\  accepted 2016

\bibitem[\protect\astroncite{{Lamb} et~al.}{2009}]{lamb09}
{Lamb}, F.~K., {Boutloukos}, S., {Van Wassenhove}, S., et~al.\  2009, ApJ, 706,
  417

\bibitem[\protect\astroncite{{Lamb} et~al.}{1977}]{lamb77}
{Lamb}, F.~K., {Fabian}, A.~C., {Pringle}, J.~E., \& {Lamb}, D.~Q.  1977, ApJ,
  217, 197

\bibitem[\protect\astroncite{{Lattimer} \& {Prakash}}{2007}]{lattimer07}
{Lattimer}, J.~M., \& {Prakash}, M.  2007, Phys.Rep, 442, 109

\bibitem[\protect\astroncite{{Lewin} et~al.}{1976}]{lewin76}
{Lewin}, W.~H.~G., {Clark}, G., \& {Doty}, J.  1976, IAU~Circular, 2922

\bibitem[\protect\astroncite{{Miller}}{2007}]{miller07}
{Miller}, J.~M.,  2007, ARA\&A, 45, 441

\bibitem[\protect\astroncite{{Miller} et~al.}{2013}]{miller13}
{Miller}, J.~M., {Parker}, M.~L., {Fuerst}, F., et~al.\  2013, ApJ, 779, L2

\bibitem[\protect\astroncite{{Mondal} et~al.}{2016}]{mondal16}
{Mondal}, A.~S., {Pahari}, M., {Dewangan}, G.~C., {Misra}, R., \&
  {Raychaudhuri}, B.  2016, submitted to MNRAS (arxiv:1604.04366)

\bibitem[\protect\astroncite{{Narita} et~al.}{2001}]{narita01}
{Narita}, T., {Grindlay}, J.~E., \& {Barret}, D.  2001, ApJ, 547, 420

\bibitem[\protect\astroncite{{Ng} et~al.}{2010}]{ng10}
{Ng}, C., {D{\i}az Trigo}, M., {Cadolle Bel}, M., \& {Migliari}, S.  2010,
  A\&A, 522, 25

\bibitem[\protect\astroncite{{Papitto} et~al.}{2013}]{papitto13}
{Papitto}, A., {D'A{\'i}}, A., {Di Salvo}, T., et~al.\  2013, MNRAS, 459, 3411

\bibitem[\protect\astroncite{{Piraino} et~al.}{2000}]{piraino00}
{Piraino}, S., {Santangelo}, A., \& {Kaaret}, P.  2000, A\&A, 360, L35

\bibitem[\protect\astroncite{{Reynolds} \& {Nowak}}{2003}]{reynolds03}
{Reynolds}, C.~S., \& {Nowak}, M.~A.  2003, Phys.Rep, 377, 389

\bibitem[\protect\astroncite{{Romanova} et~al.}{2008}]{romanova08}
{Romanova}, M.~M., {Kulkarni}, A.~K., \& {Lovelace}, R.~V.~E.  2008, ApJ, 673,
  L171

\bibitem[\protect\astroncite{{Ross} \& {Fabian}}{2005}]{reflionx}
{Ross}, R.~R., \& {Fabian}, A.~C.  2005, MNRAS, 358, 211

\bibitem[\protect\astroncite{{Seifina} \& {Titarchuk}}{2011}]{seifina11}
{Seifina}, E., \& {Titarchuk}, L.  2011, ApJ, 738, 128

\bibitem[\protect\astroncite{{Strohmayer} et~al.}{1996}]{strohmayer96}
{Strohmayer}, T.~E., {Zhang}, W., {Swank}, J.~H., et~al.\  1996, ApJ, 469, L9

\bibitem[\protect\astroncite{{Tarana} et~al.}{2011}]{tarana11}
{Tarana}, A., {Belloni}, T., {Bazzano}, A., {M{\'e}ndez}, M., \& {Ubertini}, P.
   2011, MNRAS, 416, 873

\bibitem[\protect\astroncite{{van Paradijs} \&
  {McClintock}}{1994}]{vanparadijs94}
{van Paradijs}, J., \& {McClintock}, J.~E.  1994, A\&A, 290, 133

\bibitem[\protect\astroncite{{Verner} et~al.}{1996}]{vern}
{Verner}, D.~A., {Ferland}, G.~J., {Korista}, K.~T., \& {Yakovlev}, D.~G.
  1996, ApJ, 465, 487

\bibitem[\protect\astroncite{{White} \& {Zhang}}{1997}]{white97}
{White}, N.~E., \& {Zhang}, W.  1997, ApJ, 490, L87

\bibitem[\protect\astroncite{{Wilms} et~al.}{2000}]{wilm}
{Wilms}, J., {Allen}, A., \& {McCray}, R.  2000, ApJ, 542, 914

\end{thebibliography}


\clearpage
\begin{table}
\caption{Spectral parameters with varied emissivity using the {\ttfamily reflionx} model\label{tab:emissivity}}
\begin{minipage}{\linewidth}
\begin{center}
\resizebox{\columnwidth}{!}{
\begin{tabular}{ccccccccc} \hline \hline
Model & Parameter\footnote{The errors on the parameters are 90\% confidence.} & Units & $q=1$ & $q=2$ & $q=3$ & $q=4$ & $q=5$ & $q$ broken\\ \hline\hline

constant & FPMA & -- & 1\footnote{These parameters were fixed.} & 1$^b$ & 1$^b$& 1$^b$& 1$^b$ & 1$^b$\\
 & FPMB & -- & $1.05\pm0.002$ & $1.05\pm0.002$ & $1.05\pm0.002$& $1.05\pm0.002$ & $1.05\pm0.002$ & $1.05\pm0.002$ \\
 & XRT & -- &$1.12\pm0.01$ & $1.12\pm0.01$ & $1.15\pm0.01$& $1.13\pm0.01$ & $1.13\pm0.02$ & $1.12\pm0.02$\\
tbabs & $N_{\rm H}$\footnote{The column density is calculated assuming \cite{wilm} abundances and \cite{vern} cross sections.}
           & $10^{22}$\,cm$^{-2}$ & $4.6\pm0.1$ & $4.5^{+0.1}_{-0.2}$ &$4.5\pm0.1$ &$4.4\pm0.1$ & $4.4\pm0.1$ & $4.4\pm0.1$\\
bbodyrad & $kT$   &    keV   & $1.54\pm0.02$  & $1.69^{+0.03}_{-0.09}$  & $1.53^{+0.02}_{-0.05}$  &  $1.52\pm0.04$ & $1.53\pm0.04$ & $1.49^{+0.06}_{-0.01}$\\
 & norm & $R_{\rm km}^2/D_{10\rm kpc}^2$  & $6.26^{+1.4}_{-0.5}$ & $4.2^{+0.7}_{-0.4}$ &$7.9^{+2.6}_{-0.7}$ & $8.5\pm2$ &  $7.7^{+3.3}_{-1.7}$ & $9.4^{+1.7}_{-2.1}$ \\
cutoffpl & $\Gamma$ & -- & $1.52^{+0.04}_{-0.03}$ & $1.51^{+0.02}_{0.01}$ &$1.54^{+0.04}_{-0.05}$ & $1.52^{+0.05}_{-0.02}$ &  $1.51\pm0.05$ & $1.52\pm0.04$\\
 & HighECut & keV   & $26^{+3}_{-1}$  & $26^{+2}_{-4}$ &$25\pm2$ & $25^{+2}_{-3}$ & $26^{+3}_{-2}$ & $26^{+3}_{-1}$\\
 & norm & Photons keV$^{-1}$ cm$^{-2}$s$^{-1}$ at 1keV  & $0.19^{+0.02}_{-0.04}$ & $0.19^{+0.13}_{-0.03}$ & $0.29^{+0.03}_{-0.07}$ & $0.26^{+0.04}_{-0.07}$ & $0.23^{+0.05}_{-0.07}$ & $0.20\pm0.06$ \\
kerrconv & $q_{\rm in}$ & -- & 1$^b$ & 2$^b$ & 3$^b$ & 4$^b$ & 5$^b$ & $>5.4$\\
 & $q_{\rm out}$ & -- & 1$^b$& 2$^b$ & 3$^b$ & 4$^b$ & 5$^b$ & $3^b$ \\
 & $a$ & -- & 0.15 $^b$ & 0.15$^b$ &0.15$^b$ &0.15$^b$ &0.15$^b$ & 0.15$^b$\\
 & Incl. & deg. & $35^{+13}_{-5}$ & $19^{+8}_{-4}$ &$37^{+1}_{-2}$ & $37^{+2}_{-3}$ & $39^{+2}_{-4}$ & $40^{+1}_{-3}$\\
 & $R_{\rm in}$ & ISCO & $3.2^{+345}_{-2.2}$ & $1.0^{+0.56}_{-0}$ &$1.0^{+0.77}_{-0}$ & $1.1^{+0.50}_{-0.1}$ &  $1.1^{+0.48}_{-0.1}$ & $1.2^{+0.4}_{-0.2}$\\
 & $R_{\rm out}$ & ISCO & 400$^b$ & 400$^b$ & 400$^b$ & 400$^b$ & 400$^b$ & $400^b$\\
reflionx & $\xi$ & erg cm s$^{-1}$ & $3142^{+987}_{-302}$ & $4564^{+2000}_{-1556}$ &$796^{+798}_{-229}$ & $984^{+722}_{-472}$ & $1208^{+559}_{-532}$ & $1135^{+350}_{-265}$\\
 & $A_{Fe}$ & -- & $0.22\pm0.1$ & $0.46^{-0.01}_{0.68}$ &$0.19\pm0.1$ & $0.17^{+0.06}_{-0.04}$ & $0.19^{+0.07}_{-0.08}$ & $0.14\pm0.01$\\
 & norm & $10^{-6}$ & $1.3^{+0.3}_{-0.2}$ & $0.9^{+0.3}_{-0.2}$ &$3.8^{+2.6}_{-1.4}$ & $3.6^{+2.3}_{-0.6}$ & $3.2^{+1.8}_{-0.9}$ & $4.1^{+1.3}_{-1.2}$\\
$\chi^2/\rm dof$ & -- & -- & 1431/1254 & 1427/1254 &1430/1254 & 1427/1254 & 1425/1254 & 1422/1253\\
 \ \\
 \hline
\end{tabular}
}
\end{center}
\end{minipage}
\end{table}

\begin{table}
\caption{Spectral parameters varying the column density and the energy band using {\ttfamily reflionx} \label{tab:parameters}}
\begin{minipage}{\linewidth}
\begin{center}
\resizebox{\columnwidth}{!}{
\footnotesize
\begin{tabular}{ccccccc} \hline \hline
Model & Parameter\footnote{The errors on the parameters are 90\% confidence.} & Units & \begin{tabular}{c}Model 1\\ Full band\end{tabular} & \begin{tabular}{c}Model 2\\ $N_{\rm H}$ fixed\end{tabular} & \begin{tabular}{c}Model 3\\ line only\end{tabular} & \begin{tabular}{c}Model 4\\ \emph{NuSTAR} only\end{tabular} \\ \hline\hline

constant & FPMA & -- & 1\footnote{These parameters were fixed.} & 1$^b$ & 1$^b$ & $1^b$\\
 & FPMB & -- &  $1.05\pm0.002$ & $1.05\pm0.002$ & $1.05\pm0.002$ & $1.05\pm0.002$ \\
 & XRT & -- &  $1.12\pm0.02$ & $1.15\pm0.01$ & $1.15\pm0.02$ & -- \\
tbabs & $N_{\rm H}$\footnote{The column density is calculated assuming \cite{wilm} abundances and \cite{vern} cross sections.}
     & $10^{22}$\,cm$^{-2}$ & $4.5\pm0.1$ & $2.9^b$  & $4.0\pm0.1$ & $2.9^b$ \\
bbodyrad & $kT$   &  keV  & $1.53^{+0.02}_{-0.05}$  &  $1.42\pm0.01$ & $1.47\pm0.01$ & $1.45^{+0.06}_{-0.03}$  \\
 & norm & $R_{\rm km}^2/D_{10\rm kpc}^2$   & $7.9^{+2.6}_{-0.7}$ & $20.6^{+0.8}_{-0.4}$ & $10.7^{+0.5}_{-0.6}$ &$10.8\pm1.8$   \\
cutoffpl & $\Gamma$ & -- &$1.54^{+0.04}_{-0.05}$  & 1.0$^{+0.02}_{-1.0}$ &  $1.34^{+0.04}_{-0.02}$ & $1.69\pm0.08$ \\
 & HighECut & keV   & $25\pm2$ & $18.0^{+0.6}_{-0.1}$ &$19.1^{+0.7}_{-0.6}$  &  $36^{+6}_{-5}$ \\
 & norm & Photons keV$^{-1}$ cm$^{-2}$s$^{-1}$ at 1keV &$0.29^{+0.03}_{-0.07}$ &  $0.065^{+0.008}_{-0.005}$ & $ 0.29\pm0.02 $  & $0.15^{+0.08}_{-0.05}$\\
kerrconv & $q$ & -- & $3^b$ & 3$^b$ & -- & $3^b$\\
 & $a$ & -- & 0.15$^b$ & 0.15$^b$ & -- & $0.15^b$ \\
 & Incl. & deg.  &$37^{+1}_{-2}$ &  $29^{+3}_{-2}$ & -- & $33^{+3}_{-2}$ \\
 & $R_{\rm in}$ & ISCO & $1.0^{+0.77}_{-0}$ &  $1.0^{+0.44}_{-0}$ & -- & $1.7^{+1.0}_{-0.7}$ \\
 & $R_{\rm out}$ & ISCO & $400^b$ &  400$^b$ & -- & $400^b$ \\
reflionx & $\xi$ & erg cm s$^{-1}$ & $796^{+798}_{-229}$ &  995$^{+30}_{-62}$ & -- & $909^{+337}_{-189}$ \\
 & $A_{Fe}$ & -- & $0.19\pm0.1$ &  0.09$\pm0.01$ & --  & $0.07^{+0.07}_{-0.01}$ \\
 & norm & $10^{-6}$  & $3.8^{+2.6}_{-1.4}$ & 3.3$^{+0.4}_{-0.2}$ & -- & $8.8^{+4.2}_{-5.6}$ \\
relline & lineE & keV & -- & -- & $7.1\pm0.1$ & -- \\
 & $q$ & -- & -- & -- & $3^b$ & -- \\
 & $a$ & -- & -- & -- & $0.15^b$ & -- \\
 & Incl. & deg & -- & -- & $18^{+5}_{-3}$ & -- \\
 & $R_{\rm in}$ & ISCO & -- & -- & $1.1^{+0.2}_{-0}$ & -- \\
 & $R_{\rm out}$ & $R_{\rm g}$ & -- & -- & $400^b$  & -- \\
 & norm & $10^{-3}$ & -- & -- &  $1.8\pm0.2$  & -- \\
$\Omega/2\pi$ & -- & -- & $0.43$ & $0.76$ & -- & -- \\
$\chi^2/\rm dof$ & -- & -- & 1430/1254 & 1712/1255 & 1532/1255 & 1256/1096 \\
 \ \\
 \hline
\end{tabular}
}
\end{center}
\end{minipage}
\end{table}

\begin{table}
\caption{Spectral Parameters using {\ttfamily relxill} \label{tab:relxill}}
\begin{minipage}{\linewidth}
\begin{center}
\footnotesize
\begin{tabular}{cccc} \hline \hline
Model & Parameter\footnote{The errors on the parameters are 90\% confidence} & Units & Value\\ \hline \hline
constant & FPMA & -- & 1\footnote{These parameters were fixed} \\
 & FPMB & -- &  $1.05\pm0.002$   \\
 & XRT & -- &   $1.15\pm0.02$  \\
tbabs & $N_{\rm H}\footnote{The column density is calculated assuming \cite{wilm} abundances and \cite{vern} cross sections.}$ & $10^{22}$cm$^{-2}$ & $3.9^{+0.4}_{-0.3}$  \\
bbody & $kT$ & keV & $0.49^{+0.06}_{-0.05}$ \\
 & norm & $L_{10^{39}\rm ergs/s}/D_{10\rm kpc}^2$ & $0.007^{+0.002}_{-0.001}$ \\
comptt & $kT_{\rm seed}$ & keV & $1.13^{+0.05}_{-0.03}$ \\
 & $kT_{e}$ & keV &  $15^{+5}_{-3}$ \\
 & optical depth & -- & $1.6^{+0.3}_{-0.6}$ \\
 & norm & -- & $0.022^{+0.002}_{-0.007}$ \\
relxill & $q_{\rm in}$ & -- & $>4$ \\
 & $q_{\rm out}$ & -- & $3^b$ \\
 & $a$ & -- & $0.15^b$ \\
 & Incl. & deg. & $37^{+6}_{-2}$ \\
 & $R_{\rm in}$ & ISCO & $1.6\pm0.4$ \\
 & $R_{\rm out}$ & $R_{\rm g}$ & $400^b$ \\
 & log$\xi$ & -- & $3.0^{+0.1}_{-2.1}$ \\
 & $A_{\rm Fe}$ & -- & $0.8\pm0.1$ \\
 & norm & $10^{-3}$ & $1.2^{+1}_{-0.2}$ \\
$\chi^2/\rm dof$ &-- & --& 1412/1250 \\
 \ \\
 \hline
\end{tabular}
\end{center}
\end{minipage}
\end{table}

\end{document}